\documentstyle[prl,aps,psfig]{revtex}
\def\Bf#1{\mbox{\boldmath{$#1$}}}
\begin{document} 
\draft
\twocolumn[\hsize\textwidth\columnwidth\hsize\csname
@twocolumnfalse\endcsname

\title{On the response of composite fermions to weak 
electrostatic potentials } 

\author{Behnam Farid}

\address{Cavendish Laboratory, Department of Physics, 
University of Cambridge,\\
Madingley Road, Cambridge CB3 0HE, United Kingdom}
%\thanks{Electronic address: bf10000@phy.cam.ac.uk}\\ 
%%and Max-Planck-Institut f\"ur Festk\"orperforschung,
%%Heisenbergstra\ss e 1,\\ 
%%70569 Stuttgart, Federal Republic of Germany } 
%\thanks{Electronic address: farid@audrey.mpi-stuttgart.mpg.de} }

%\date{... }
\date{\today}

\maketitle

\begin{abstract}
\leftskip 54.8pt
\rightskip 54.8pt
We establish that the response to static perturbations of two-dimensional 
electron systems (2DESs), exposed to magnetic field ${\Bf B}$, in states 
with the Landau-level filling fractions $\nu_{\rm e}$ close to one such as 
$1/2$, is singularly different from that of 2DESs at small $\|{\Bf B}\|$. 
Thus in addition to demonstrating the sever inadequacy of the treatment of 
composite fermions as non-interacting particles, we show, in contradiction 
to a recent theoretical finding, that the physical origin of the observed 
behaviour of the dc longitudinal magneto-resistivity for $\nu_{\rm e} 
\approx 1/2$ in periodically modulated 2DESs remains to be determined.
\end{abstract}

 ]
\narrowtext
%\widetext

Advance towards understanding of the fractional quantum Hall effect 
(FQHE) was made through the variational wavefunction due to Laughlin 
\cite{Laughlin} for the corresponding states. Subsequent progress was 
made through the application of a statistical transformation 
\cite{Wilczek}, transmuting electrons into hard-core bosons, by Girvin 
and MacDonald \cite{Girvin} who showed that the equal-time boson-boson 
correlation function of the transformed Laughlin wavefunction exhibits 
algebraic off-diagonal long-range order. This statistical transformation 
has been an essential ingredient in considerations by the latter authors 
\cite{Girvin}, Read \cite{Read} and Zhang, Hansson and Kivelson 
\cite{Zhang} in formulating a Ginzburg-Landau theory for the FQHE. The 
microscopic formulation of this theory relies on a statistical gauge 
field associated with the Chern-Simons (CS) action in $2+1$ dimensions 
whose coupling constant $\Theta_{\rm CS}$ is equal to an odd multiple of 
$2\pi$. 

An alternative approach to the FQHE was advanced by Jain \cite{Jain}, 
who through introducing a variational wavefunction, involving the Laughlin 
wavefunction pertaining to a filled Landau level (LL) to whose constituent 
particles an {\sl even} number of flux quanta are attached, ascribed the 
phenomenon of FQHE to the integer QHE (IQHE) of `composite fermions' (CFs). 
A CF is thus composed of an electron and $2 n$, $n=1,2,\dots$, quantised 
magnetic flux tubes of magnetic flux $\Phi_0 \equiv h/e$ (throughout, 
$-e < 0$). The total magnetic flux due to these flux tubes results in a 
reduced effective magnetic flux density $\Delta {\Bf B}$ of whose 
corresponding LLs (for CFs) in the case where the {\sl electronic} 
$\nu_{\rm e}$ is equal to $p/(2 n p + 1)$, exactly $p$ levels are 
completely filled {\sl and} none is fractionally occupied. We note in 
passing that for $\nu_{\rm e} \in (0,1]$, the `particle-hole' symmetry 
(which is an exact property for the case where the lowest LL is decoupled 
from the higher LLs) implies that $\nu_{\rm e}' {:=} 1-\nu_{\rm e}$ 
corresponds to a FQH state if $\nu_{\rm e}$ does to one, so that the Jain 
series $\nu_{\rm e} = p/(2 n p + 1)$ implies FQH states at $\nu_{\rm e}' 
= 1 -p/(2 n p + 1)$; each of these, on account of the `law of 
corresponding states' \cite{Kivelson}, in turn implies FQH states at 
$\nu_{\rm e}'' = m + \nu_{\rm e}$, where $m$ is an integer. For increasing 
values of $p$, $\nu_{\rm e} = p/(2 n p + 1)$ and $\nu_{\rm e}' {:=} 
1-\nu_{\rm e}$ approach $\nu_{\rm e}=1/(2 n)$ from left and right, 
respectively. The filling fraction $\nu_{\rm e}=1/(2n)$, although a limit 
of those associated with FQH states for $p\to\infty$, does {\sl not} 
itself correspond to a QH state, in that its associated state is 
compressible \cite{Halperin}. The notable fractions in this category 
are $\nu_{\rm e}=1/2, 1/4$ and $\nu_{\rm e}' = 3/4$. 

The scenario as put forward by Jain \cite{Jain} finds its microscopic 
justification in the fermionic version of the CS field theory in $2+1$ 
dimensions where $\Theta_{\rm CS} = 2 n \times 2\pi$, $n=1,2,\dots$. 
Indeed, from the mean-field Hamiltonian of this theory for {\sl extended} 
2DESs \cite{Dai},\cite{Chen},\cite{Halperin}, it is apparent that CFs 
are exposed to an effective magnetic flux density equal to $\Delta B$ 
(here and further on, $B$ and $\Delta B$ denote the amplitudes of ${\Bf 
B}$ and $\Delta {\Bf B}$ which are assumed to be perpendicular to the 
plane of 2DESs). A most appealing aspect of the fermionic CS field theory 
in $2+1$ dimensions was exposed by Lopez and Fradkin \cite{Lopez} who 
demonstrated that all physical amplitudes calculated within this theory 
are identical with those calculated within the conventional theory where 
$\Theta_{\rm CS} =0$. It follows that in cases where $\Theta_{\rm CS} = 
2 n \times 2 \pi$, the CS field theory has the {\sl practical} advantage 
that its mean-field approximation captures the essential aspects of the 
FQH states, such as their incompressibility. CFs have found experimental 
support in the surface-acoustic-wave resonance experiments by Willett, 
{\sl et al.} \cite{Willett1}, geometric resonance measurements by Kang, 
{\sl et al.} \cite{Kang} and Willett,{\sl et al.} \cite{Willett2} and 
magnetic focusing experiments by Goldman, {\sl et al.} \cite{Goldman} 
and Smet, {\sl et al.} \cite{Smet1}.

A recent experimental work by Smet, {\sl et al.} \cite{Smet2} on 
periodically and supposedly {\sl weakly} modulated 2DESs has revealed 
some unexpected aspects concerning the dc longitudinal magneto-resistivity 
$\varrho_{xx}$ as function of $B$ close to $\nu_{\rm e}=1/2$. These include 
the observation that a supposedly $1$\% modulation amplitude in the number 
density $n_e$ of electrons gave rise to a positive magneto-resistance, 
for $\vert B - B_{1/2}\vert$, typical of a charge-density modulation 
amplitude of $10$\% (see below, in particular the paragraph containing
the definition for $\epsilon$); here $B_{\nu_{\rm e}} \equiv \Phi_0 
n_e/\nu_{\rm e}$ is the magnetic flux density corresponding to 
$\nu_{\rm e}$. In this connection, the authors \cite{Smet2} state 
``Clearly some ingredients are missing to form a consistent model, 
unless our implicit assumption, that the strength of the density 
perturbation taken from low field data can be transferred to the high 
field regime, is incorrect.'' One of the findings of the present work 
is that this is indeed the case. 

The observations by Smet, {\sl et al.} \cite{Smet2}, have been the subject 
of theoretical consideration by von Oppen, {\sl et al.} \cite{vonOppen}, 
Mirlin, {\sl et al.} \cite{Mirlin}, and Zwerschke and Gerhardts 
\cite{Zwerschke}. By taking into account the differences in the scattering 
properties of random electrostatic and magnetic fields (in the case at hand 
the two are related via the relationship $\Delta B \equiv \Phi_0 n_e/p$), 
Mirlin, {\sl et al.} \cite{Mirlin} and Zwerschke and Gerhardts 
\cite{Zwerschke} obtain results for $\varrho_{xx}$, concerning a 
periodically modulated 2DES along the $x$ direction, which are in 
reasonable agreement with experimental results. By further solving a 
linearised Boltzmann equation for CFs which in a self-consistent manner 
takes account of the electric field generated by the motion of the flux 
tubes associated with CFs, the latter authors obtain results which are 
in excellent agreement with experiment. A crucial observation by 
Zwerschke and Gerhardts \cite{Zwerschke} is that the so-called `channeled 
orbits', which are neglected in the approximate treatment of the Boltzmann 
equation due to Beenakker \cite{Beenakker}, determine the specific v-shape 
in the experimentally observed $\varrho_{xx}(B)$ close to $B = B_{1/2}$.

Since in the above-cited works the CFs are considered as non-interacting 
particles, the strength of the modulating potential as `seen' by CFs is 
not the bare one, namely $v_{\rm ext}(x)$, but one which is screened,
namely $v_{\rm eff}(x)$. This becomes evident by applying the mean-field 
approximation subsequent to having introduced $v_{\rm ext}(x)$ into the 
many-body Hamiltonian of an otherwise uniform system. Assuming
$v_{\rm ext}(x) = v_{Q} \cos(Q x)$, with $Q=2\pi/a_x$, {\sl linear} 
screening suffices for small $v_Q$ and one has $v_{\rm eff}(x) 
= \big(\varepsilon_{\rm pp}^{-1}(Q,0) v_Q\big) \cos(Q x)$, where 
$\varepsilon_{\rm pp}^{-1}(Q,0)$ stands for the inverse of the static 
`proton-proton' dielectric function \cite{Heine}; 
$\varepsilon_{\rm pp}^{-1}(q,0) = 1 + \big(v(q)+\kappa_{\rm xc}(q)\big)
\chi(q,0)$ where $\chi$ denotes the density-density response function 
of the interacting system and $v(q)$ the Fourier transform of the 
electron-electron interaction function which in this work we assume
to be the Coulomb interaction function so that $v(q)=e^2/(2\epsilon_0
\epsilon_r q)$ with $\epsilon_0$ the vacuum permittivity and $\epsilon_r$ 
for the relative dielectric constant; $\kappa_{\rm xc} (q)$ is the 
exchange and correlation contribution to the interaction. For the 
$\varepsilon^{-1}$ specific to screening of the electron-electron 
interaction, one has $\varepsilon^{-1}(q,0) = 1 + v(q) \chi(q,0)$. Since 
$\kappa_{\rm xc}(q)/v(q) = {\cal O}(q)$ for $q\to 0$ and since in our 
considerations $Q \ll k_F$ (we consider $n_e = 1.36 \times 
10^{11}$~cm$^{-2}$ and $a_x = 400$~nm), with ${\Bf k}_F$ the Fermi 
wavevector, we can neglect $\kappa_{\rm xc}(q)$ and thus identify 
$\varepsilon_{\rm pp}^{-1}$ with $\varepsilon^{-1}$. 

A modulation is considered weak if the dimensionless quantity
$\epsilon {:=} e \varepsilon_{\rm pp}^{-1}(Q,0) v_Q/(\sqrt{2} E_F)$,
is small ($\epsilon$ as defined here is identical with that by Beenakker 
\cite{Beenakker} and is $1/\sqrt{2}$ times the $\epsilon$ by Zwerschke and 
Gerhardts \cite{Zwerschke} which coincides with $\eta$ by Mirlin, {\sl 
et al.} \cite{Mirlin}). This is because according to the Thomas-Fermi 
approximation (TFA), $\varepsilon^{-1}_{\rm TF}(q,0) = 1/(1 + g_e(E_F) 
v(q))$ where $g_e(\omega) = m_b/(\pi\hbar^2)$ is the {\sl total} density 
of one-electron states for spin-unpolarised 2DES, with $m_b$ the bare 
band-electron mass; with $v(q)$ the Coulomb interaction, $g_e(E_F) v(q) 
= a_0 q/2$, where $a_0 {:=} 4\pi\epsilon_0 \epsilon_r \hbar^2/(m_b e^2)$ 
is the effective Bohr radius (in this work, $a_0 = 9.79 \times 10^{-9}$~m), 
so that with $a_x \approx 400 \times 10^{-9}$~m, $\varepsilon^{-1}_{\rm TF}
(Q,0) \approx a_0 Q/2 \ll 1$ {\sl and} $\chi_{\rm TF}(Q,0) \approx -1/v(Q)$ 
from which, making use of $E_F g_e = n_e$, one obtains $\delta n_e(Q)/n_e 
\equiv e \chi(Q,0) v_Q/n_e \approx \sqrt{2} \epsilon$. By {\sl assuming} the
validity of the TFA for CFs, one has $\epsilon_{\rm CF} \approx \epsilon$ 
(with $m_{\star}\sim 5 m_b$ the CF effective mass, $\varepsilon^{-1}_{\rm 
TF}(Q,0) \approx (2 m_b/m_{\star}) a_0 Q/2 \ll 1$, so that for CFs also 
$\chi_{\rm TF}(Q,0) \approx -1/v(Q)$). Thus the {\sl assumption} with 
regard to the validity of the TFA for CFs implies that a `weak' 
perturbation for electrons is equally `weak' for CFs and vice versa. It 
is common practice to assign a fixed, i.e. independent of $B$, value to 
$\epsilon$, reflecting the constancy of $\epsilon$ within the TFA. The 
significance of $\epsilon$ is clarified by considering the expression 
for $\varrho_{xx}$ as determined by Beenakker (Eq.~(6) in 
Ref.~\cite{Beenakker}); by choosing $\epsilon =0.015$ \cite{Beenakker}, 
this $\varrho_{xx}$ accurately describes the magneto-resistance oscillations 
as observed by Weiss, {\sl et al.} \cite{Weiss}.

For a 2DES in the paramagnetic phase exposed to magnetic field, we have  
$\chi(q,\omega) = \chi^0(q,\omega)/\big(1- (v(q)+\kappa_{\rm xc}(q))
\chi^0(q,\omega)\big)$, where \cite{Glasser}
\begin{equation}
\label{e1}
\chi^0(q,0) = \frac{- m_b}{\hbar^2\pi^2}
{\cal F}_{[E_F/(\hbar\omega_c)]}
\left(\frac{\hbar q^2}{2 m_b \omega_c}\right);
\end{equation}
here $[x]$ denotes the greatest integer less than or equal to $x$ and 
${\cal F}_N(a) {:=} \int_0^{\pi} {\rm d}\tau\; \tau \sin(a\sin\tau) 
\exp(-a[1+\cos\tau]) G_N(2a [1+\cos\tau])$; $G_N(x)\equiv 1$, when $N=0$, and 
$G_N(x)\equiv  L_N(x) + 2 L_{N-1}^{(1)}(x)$, when $N\geq 1$. Here $L_N(x)$ 
denotes the $N$th-order Laguerre polynomial and $L_N^{(m)}(x) {:=} (-1)^m 
{\rm d}^m L_{N+m}(x)/{\rm d} x^m$. By neglecting $\kappa_{\rm xc}(q)$, 
the expression for $\chi(q,0)$ reduces to the random-phase approximation
(RPA). Since in our applications in the present work $Q \ll k_F$, we 
neglect $\kappa_{\rm xc}$ and thus employ the RPA for $\chi(q,0)$. 

For CFs, the counterpart of $\chi(q,\omega)$ is denoted by $-K_{00}(q,
\omega)$ which is the $(\mu,\nu)=(0,0)$ component of a $2\times 2$
matrix ${\bf K}(q,\omega)$ with $K_{\mu,\nu} \equiv ({\bf K})_{\mu,
\nu}$. Here $\mu,\nu\in \{0,1\}$, with $0$ indicating the `time'
component and $1$ the transverse `space' component \cite{Halperin}.
Within the RPA, we have ${\bf K} = {\bf K}^0 \big( {\bf I} + {\bf U} 
{\bf K}^0\big)^{-1}$, where ${\bf U} {:=} {\bf V} + {\bf C}^{-1}$ in 
which $V_{00} = v(q)$ and $V_{\mu,\nu} \equiv 0$ for $(\mu,\nu)\not=
(0,0)$, and for $\nu_{\rm e} = p/(2 n p + 1)$, $C_{00}\equiv C_{11} 
\equiv 0$, $C_{01} \equiv - C_{10} \equiv i q/(h {\tilde\phi})$, where 
${\tilde\phi} {:=} 2 n$ \cite{Halperin}, \cite{Dai}, \cite{Simon}. It 
can be shown \cite{Simon} that ${\bf K}^0 = {\bf T}^{-1} {\bf S} 
{\bf T}^{-1}$, where $T_{00} \equiv i (e/q) \sqrt{i\omega}$, $T_{11} \equiv 
e/\sqrt{i\omega}$, $T_{01}\equiv T_{10} \equiv 0$ and
$S_{00} \equiv i p e^2 (\omega/\Delta\omega_c) \Sigma_0/h$,
$S_{11} \equiv i p e^2 (\Delta \omega_c/\omega) (\Sigma_2+1)/h$ and
$S_{10} \equiv -S_{01} \equiv p e^2 \Sigma_1/h$, in which 
$\Delta\omega_c {:=} e \Delta B/m_b$; for $\omega=0$ and $p\geq 1$,
\begin{eqnarray}
\label{e2}
& &\Sigma_j \equiv
\frac{-\exp(-Y)}{p} \sum_{\ell=0}^{p-1}
\sum_{m=p}^{\infty}\nonumber\\
& &\big\{\frac{\ell!}{(m-\ell) m!} Y^{m-\ell-1}
\{ L_{\ell}^{(m-\ell)}(Y)\}^{2-j}\nonumber\\
& &\times\{ (m-\ell-Y) L_{\ell}^{(m-\ell)}(Y) 
+ 2 Y {\rm d} L_{\ell}^{(m-\ell)}(Y)/{\rm d} Y \}^j\big\},
\end{eqnarray}
where $Y {:=} (q R_{\Delta})^2/(4 p)$, with $R_{\Delta} \equiv \hbar 
k_F/(e \Delta B)$, the effective semi-classical cyclotron radius for 
CFs. Since the state of CFs corresponds to that of fully-polarised 
electrons, $k_F = \sqrt{4\pi n_e}$. One obtains
\begin{equation}
\label{e3}
K_{00}(q,\omega=0) = \frac{-p q^2 \Sigma_0}{h\Delta\omega_c 
\Upsilon},
\end{equation}
where $\Upsilon {:=} (1 + p {\tilde\phi}\Sigma_1)^2 - p q^2 v(q) 
\Sigma_0/(h \Delta\omega_c) - p^2 {\tilde\phi}^2 \Sigma_0 (\Sigma_2+1)$.
Assuming the CFs to have the mass $m_{\star}$, a direct change of $m_b$ 
into $m_{\star}$, or of $\Delta\omega_c$ into $\Delta\omega_{c\star} 
{:=} e \Delta B/m_{\star}$, in the above expressions, would result in 
a $K_{00}(q,\omega)$ that violates a requirement implied by a theorem 
due to Kohn \cite{Kohn}, \cite{Zhang}, \cite{Halperin}. Simon and 
Halperin \cite{Simon} 
have put forward modifications, denoted by MRPA (`modified RPA') and 
M$^2$RPA, through which any desired mass can be assigned to CFs without 
ill consequences of the kind. It turns out that M$^2$RPA is superior to 
MRPA, however they are identical in so far as $K_{00}(q,\omega)$ is 
concerned. We indicate the MRPA for ${\bf K}$ by $\widetilde{{\bf K}}$. 
With ${\bf K}_{\star}^0$ denoting the ${\bf K}^0$ in which 
$\Delta\omega_c$ is replaced by $\Delta\omega_{c\star}$, we have 
$\widetilde{{\bf K}} = {\bf K}^0_{\star}\big({\bf I} + \{ {\bf U} + 
{\bf F}\} {\bf K}^0_{\star}\big)^{-1}$, where $F_{00} \equiv 
(m_{\star}-m_b) \omega^2/(n_e q^2)$, $F_{11} \equiv -(m_{\star}-m_b)/n_e$, 
$F_{01}\equiv F_{10} \equiv 0$ \cite{Simon}. A straightforward 
calculation yields
\begin{eqnarray}
\label{e4}
& &\widetilde{K}_{00}(q,0) = K_{\star 00}(q,0) 
-\frac{{\cal N}_1 {\cal N}_2}{{\cal D}},\\
\label{e5}
{\cal N}_1 &\equiv& \Big(\frac{p q}{h \Upsilon_{\star}}\Big)^2
\frac{m_{\star} - m_b}{n_e}\nonumber\\
& &\;\;\times \{(1+ p {\tilde\phi}\Sigma_1)\Sigma_1 - p {\tilde\phi}
\Sigma_0 (\Sigma_2+1)\},\\ 
\label{e6}
{\cal N}_2 &\equiv& 
\frac{p q}{\Delta\omega_{c\star}} \Sigma_0 \{\frac{q v(q)}{h}
\Sigma_1 -\frac{{\tilde\phi}}{q}\Delta\omega_{c\star} 
(\Sigma_2+1)\}\nonumber\\
& &+\Sigma_1 \{1 -\frac{p q^2 v(q)}{h\Delta\omega_{c\star}} 
\Sigma_0 + p {\tilde\phi}\Sigma_1\},\\
{\cal D} &\equiv& 1 + \frac{(m_{\star} - m_b) p}{h n_e \Upsilon_{\star}}
\nonumber\\
& &\times \big\{ p q \Sigma_1 \{\frac{q v(q)}{h}\Sigma_1
-\frac{{\tilde\phi}}{q} \Delta\omega_{c\star} (\Sigma_2+1)\}
\nonumber\\
& &\;\;\; +
\Delta\omega_{c\star} (\Sigma_2+1)
\{1 -\frac{p q^2 v(q)}{h\Delta\omega_{c\star}}\Sigma_0
+p {\tilde\phi}\Sigma_1\}\big\}.
\end{eqnarray}
Above $\Upsilon_{\star}$ is the counterpart of $\Upsilon$.

In our calculations concerning CFs, we employ $\nu_{\rm e} = h n_e/(e B)$ 
and subsequently evaluate $p$ according to $p = [\nu_{\rm e}/(1-2 n 
\nu_{\rm e})]$ if $\nu_{\rm e} < 1/(2 n)$, and $p = [(1-\nu_{\rm e})/(1- 
2 n (1-\nu_{\rm e}))]$, if $\nu_{\rm e} > 1/(2 n)$ (in the present work, 
$n=1$); for $[x]$ see text following Eq.~(\ref{e1}) above. This procedure 
introduces some error, that is in general $p/(2n p + 1)$, or $1 - p/(2n p 
+1)$ if $\nu_{\rm e} > 1/(2n)$, determined from the thus-obtained $p$ is 
not exactly equal to $\nu_{\rm e} = h n_e/(e B)$. However, explicit 
calculation shows that for the range of $B$ considered in this work, the 
relative error in $\nu_{\rm e}$ (which oscillates around zero and whose 
magnitude increases with increasing $\vert\Delta B\vert$) never exceeds 
$0.06$\%.
% 1. [29 lines]
%\pagebreak
%\clearpage
\begin{figure}[t!]
\protect
\label{fi1}
\centerline{
\psfig{figure=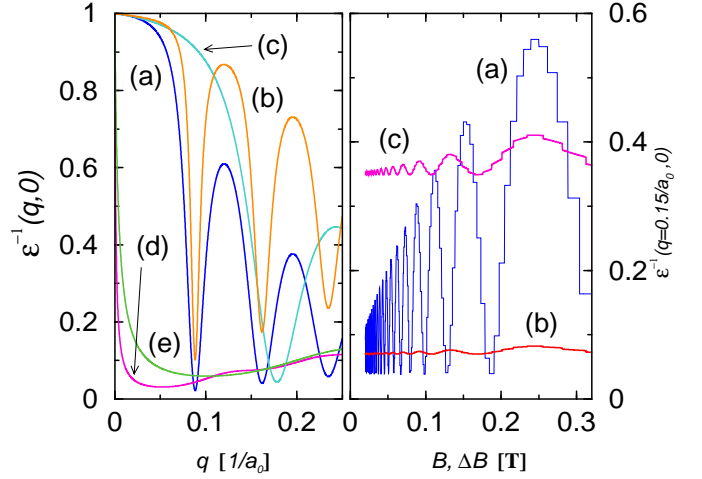,width=3.5in} }
\vskip 20pt
\caption{\rm (colour)
{\it Left panel}:
$\varepsilon^{-1}(q,0)$ for CFs ((a), (b) and (c)) and electrons ((d) 
and (e)); (a) and (b) ((d)) correspond to $\Delta B (B) =0.2$~T while (c) 
((e)) corresponds to $\Delta B (B) =0.4$~T. (a) and (c) are based on 
$\widetilde{\bf K}$ with $m_{\star} = 5\times m_b$, while for curve (b) 
$m_{\star} = m_b \equiv 0.067 m_e$, where $m_e$ is the electron mass in 
vacuum; the coincidence of the tangents of curves (a) and (b) in the 
neighbourhood of $q=0$ indicates that $\widetilde{K}_{00}(q,0)$ is indeed 
in agreement with the requirement of the Kohn theorem. Here $a_0 = 
9.79$~nm, corresponding to $\epsilon_r =12.4$ and $m_b$. All results 
concern $n_e = 1.36 \times 10^{11}$~cm$^{-2}$ so that $k_F^{\rm elec.} 
= 0.91/a_0$, $k_F^{\rm CF} = 1.28/a_0$ ($\equiv\sqrt{2} k_F^{\rm elec.}$).
{\it Right panel}:
$\varepsilon^{-1}(q,0)$ as a function of $\Delta B$ (for CFs with 
$m_{\star} = 5\times m_b$ --- curve (a)) and $B$ (for electrons --- 
curve (b)) at $q=0.15/a_0$; this $q$ coincides with $Q = 2\pi/a_x$ 
corresponding to $a_x \approx 400$~nm. Curve (c) is $5 \times$~(b). }
\end{figure}
In Figs.~1 and 2 we present results of our numerical calculations. The 
sever inadequacy of the TFA in particular for CFs is apparent. One 
observes that over considerable range of values of $\Delta B$, 
$\varepsilon^{-1}(q,0)$, and therefore $\epsilon$ defined above,
pertaining to CFs is by one order of magnitude larger than that for 
electrons. This observation is in line with the statement by Smet, 
{\sl et al.} \cite{Smet2} quoted above. One further observes that in 
the case of CFs, $\epsilon$ strongly depends on $\Delta B$. The 
similarities between $-K_{\star\, 00}^0(q,0) \equiv p q^2\Sigma_0/(h \Delta
\omega_{c\star})$ and $\chi^0(q,0)$, which describe the number-density 
response to the {\sl total} (i.e. external plus induced) potential for systems 
of CFs and electrons, respectively (Fig.~2, right panel), and lack
hereof in the corresponding $\varepsilon^{-1}(q,0)$, is indicative of the
significance of $\Sigma_1$ and $\Sigma_2$ which have no counterparts in 
the case of electrons in weak magnetic fields. We note in passing that 
for $\Delta B$ and $B$ approaching zero, the oscillations of $-K_{\star\, 
00}^0(q,0)$ and $\chi^0(q,0)$ shift towards $q=0$, so much so that in 
the limits $\Delta B = 0$ and $B = 0$ all oscillations collapse at $q=0$ 
and $-K_{\star\, 00}^0(q\to 0,0)$ and $\chi^0(q\to 0,0)$ take on 
non-vanishing constant values as expected from compressible liquids. As 
is clearly seen from the left panels of Figs.~1 and 2, for decreasing
$B$ and $\Delta B$ the functions $\varepsilon^{-1}(q,0)$ are increasingly 
`pushed' towards lower values of $q$, signalling the emergence of the 
mentioned compressible states at $B=0$ and $\Delta B=0$. We point out 
that the step-like behaviour in $\varepsilon^{-1}(q,0)$ in the right panel 
of Fig.~1 is associated with the well-known cusps in the total energies as 
functions of magnetic flux density of respective systems, mathematically 
most clearly evident from $[E_F/(\hbar\omega_c)]$, which implies 
truncation, in the expression for $\chi^0(q,0)$ in Eq.~(\ref{e1}); the 
{\sl flatness} of the plateaus of $\varepsilon^{-1}(q,0)$ corresponding 
to CFs, on the other hand, is an artifact of the error associated with 
our restriction of $\nu_{\rm e}$ to forms $p/(2 n p + 1)$ and 
$1-p/(2n p + 1)$.
% 2. [22 lines]
%\pagebreak
%\clearpage
\begin{figure}[t!]
\protect
\label{fi2}
\centerline{
\psfig{figure=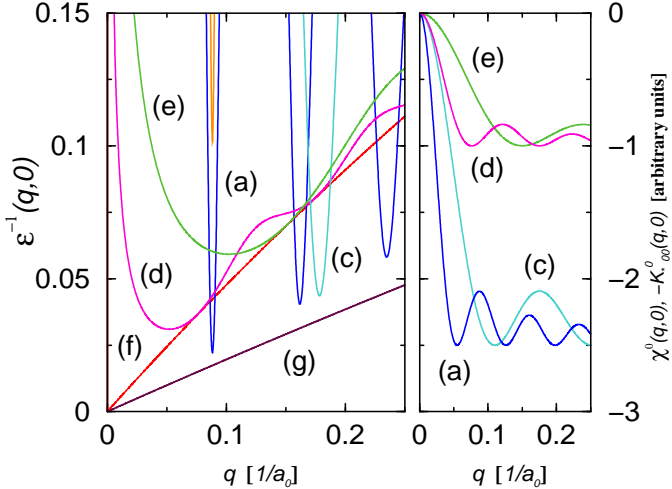,width=3.5in} }
\vskip 20pt
\caption{\rm (colour)
{\it Left panel}:
Similar to Fig.~1 but over a restricted range of values of 
$\varepsilon^{-1}(q,0)$ (the markings (a)-(e) coincide with those in 
Fig.~1). Curves (f) and (g) are the TFAs for $\varepsilon^{-1}(q,0)$ 
for electrons and CFs (using $m_{\star} = 5\times m_b$), respectively. 
{\it Right panel}:
The static density-density response functions of {\sl non}-interacting 
CFs, $-K^0_{\star\,00}(q,0)$ ((a) and (c), concerning $\Delta B=0.2, 
0.4$~T, respectively), and that of electrons, $\chi^0(q,0)$ ((d) and 
(e) --- $B=0.2, 0.4$~T). }
\end{figure}

In conclusion, by calculating the response functions of electrons and 
CFs, we have established that describing the transport properties of CFs 
in spatially modulated systems along the same lines as non-interacting 
electrons in weak magnetic fields is inadequate. The (excellent) agreement 
between the theoretical results by Mirlin, {\sl et al.} \cite{Mirlin}
and Zwerschke and Gerhardts \cite{Zwerschke} with experiments are 
therefore to be attributed to the freedom in the choice of the 
parameter(s) of the collision operator in the Boltzmann equation dealt 
with by the authors. We are presently investigating the transport 
properties of CFs along the lines as put forward by Kim, {\sl et al.} 
\cite{Kim}.

\vspace{0.2cm}
It is a pleasure for me to thank Professor Peter B. Littlewood for
invaluable support and kind hospitality at Cavendish Laboratory. With 
appreciation I acknowledge stimulating discussions with Dr Jurgen H. 
Smet and Professor Klaus von Klitzing at the initial stages of this 
work. 

%_______________________________________________________________
% 

\end{document}